\documentclass[runningheads]{llncs}

\usepackage{amsmath}
\usepackage{graphicx}
\usepackage{booktabs}
\usepackage{xcolor}
\usepackage[hidelinks]{hyperref}
\usepackage{xurl}
\usepackage{microtype}
\usepackage{newunicodechar}
\usepackage{textcomp}
\newunicodechar{Δ}{\ensuremath{\Delta}}
\newunicodechar{α}{\ensuremath{\alpha}}
\newunicodechar{β}{\ensuremath{\beta}}
\newunicodechar{ε}{\ensuremath{\varepsilon}}
\newunicodechar{δ}{\ensuremath{\delta}}
\newunicodechar{√}{\ensuremath{\surd}}
\newunicodechar{×}{\ensuremath{\times}}
\newunicodechar{≥}{\ensuremath{\geq}}
\newunicodechar{≤}{\ensuremath{\leq}}
\newunicodechar{≈}{\ensuremath{\approx}}
\newunicodechar{∈}{\ensuremath{\in}}
\newunicodechar{→}{\ensuremath{\rightarrow}}
\newunicodechar{·}{\ensuremath{\cdot}}

\begin{document}

\title{Test-Input Generation for Tensor Programs:\\What Actually Finds Kernel Bugs}
\titlerunning{Test-Input Generation for Tensor Programs}

\author{Dipankar Sarkar\orcidID{0000-0001-5431-6367}}
\authorrunning{D. Sarkar}
\institute{Arizona State University, USA \\
\email{dsarkar3@asu.edu}}

\maketitle

\begin{abstract}
Test-input generation for tensor kernels is folkloric. Most projects
pick a representative shape and dtype, run a fixed-shape
allclose-style check, and ship. We make the choices explicit and
measure them. Using the gpuemu op-schema-aware seeded
fuzzer~\cite{gpuemuP1}, we evaluate seven test-generation strategies
across a 26-op corpus (16 correct controls and 10 LLM-style buggy
variants seeded with documented transcription patterns) on an
RTX~3060 GPU instance. Strategies vary the shape candidate set, the
dtype mix, and the input value distribution. We report each strategy
on two axes: bug recall and control false-positive (FP) rate.
Boundary-only shape sampling is the operationally safe winner: 78\%
recall on the 10 buggy kernels with 0\% FP on the 16 controls.
Adversarial value sampling reaches higher recall (99\%) but inflates
control FP to 94\% because the strategy injects NaN and Inf inputs
and the validator's NaN check fires on every kernel that propagates
them, not only on buggy kernels. On the two softmax tail-mask bugs the ``regular'' strategy (no
boundary shapes) catches 0\%, while boundary raises recall to 100\%
and 62\% respectively. That gap is the clearest single signal in the
data. The corpus result is about which seeded bug patterns each
strategy catches, not about the bug rate of any specific deployed
LLM.
\keywords{kernel testing \and fuzzing \and boundary value analysis
\and Triton \and LLM-generated code}
\end{abstract}

\section{Introduction}
\label{sec:intro}

Generated GPU kernels from recent benchmarks
\cite{kernelbench2025,geak2025,kernelband2025,stark2025} are tested
by an oracle that takes inputs from somewhere. Where is rarely
documented and almost never compared. The ``regular shape'' default
(one shape per operator, one dtype, uniform random values) is the
convenient choice, but it has two empirically measured failure
modes.

\textbf{Shape-dependent bugs.} Tail-mask leak in reductions and
accumulator overwrite in matmul only surface at specific boundary
shapes (e.g., $H = 3$, $K = 1$). A test at $H = 256, K = 16$ cannot
see them.

\textbf{Magnitude-sensitive bugs.} Overflow with extreme values, and
special-case handling for zero, Inf, and NaN, only surface under
input distributions that include those values.

This paper measures both gaps and quantifies what each strategy buys.
The contributions are four.

\begin{enumerate}
\item A strategy taxonomy for tensor-kernel test-input generation,
exposed as switchable knobs on the gpuemu fuzzer~\cite{gpuemuP1}.
The knobs are the shape candidate set (boundary, regular, or default
mix), the dtype set, and the value distribution (Uniform,
NaNInjected, or Adversarial).
\item A controlled ablation across 26 ops, 7 strategies, and 8
iterations per (strategy, kernel), 1{,}456 cases on a single
RTX~3060.
\item A ranked table that reports each strategy on both recall (on
the 10 buggy kernels) and control FP rate (on the 16 correct
kernels). Adversarial reaches 99\% recall but 94\% FP. Boundary
reaches 78\% recall at 0\% FP. The ``regular'' strategy
loses an entire bug class (0\% recall on tail-mask bugs).
\item A measured explanation of the recall-FP trade-off for the
adversarial and NaN-injected strategies. Both inject non-finite
inputs, both inflate FP, and the inflation traces to the validator's
NaN check rather than to genuine output divergence.
\end{enumerate}

\section{Related Work}
\label{sec:related}

\textbf{Boundary-value testing.} A classical software-testing
principle dating to Myers~\cite{myers1979}. The highest defect density
sits at the boundaries of input partitions. Modern coverage-guided
fuzzers (AFL, libFuzzer) implement variants of this. For tensor
kernels the analogue is \emph{shape boundaries} (1, prime, power of
two $\pm 1$) and \emph{value boundaries} (0, subnormal, near
fp-max, $\pm$Inf, NaN).

\textbf{DL library fuzzing.} FreeFuzz~\cite{freefuzz2022},
DocTer~\cite{docter2022}, DeepREL~\cite{deeprel2022}, and
NablaFuzz~\cite{nablafuzz2023} all fuzz the API layer of TensorFlow,
PyTorch, and JAX. Their generation strategies are mostly value
mutational and they evaluate at the API level, not the kernel level.
Coverage-guided variants~\cite{flashfuzz2025} focus on syntactic
coverage of the API surface, not the operator's input domain. None of
them publish a per-strategy ablation of the kind below.

\textbf{Kernel-level metamorphic testing.} A few works derive
metamorphic relations for specific operators (e.g., softmax
shift-invariance), but the practice has not propagated to the LLM-kernel
ecosystem.

\textbf{LLM-kernel benchmarks.} The benchmarks
\cite{kernelbench2025,geak2025,kernelband2025,stark2025,kernelbenchx2026}
are usually one-shape, one-dtype. KernelBench's~\cite{kernelbench2025}
correctness oracle is \texttt{torch.allclose} on the reference shape.
Without varied inputs an entire class of bugs is invisible. This is
the gap we measure.

\section{Method}
\label{sec:method}

\subsection{Strategy parameters}
\label{sec:method:strategy}

A strategy is a triple (\texttt{op\_schema}, \texttt{dtypes},
\texttt{value\_distribution}) applied uniformly to every (op, iter)
under the strategy.

{\sloppy
\begin{description}
\item[\texttt{op\_schema}] a per-input shape generator. Each op ships a
native schema (mixed boundary and regular candidates). A strategy can
override the candidate set for any dim. For example \texttt{boundary}
restricts $H$ to $\{1, 3, 7\}$ and \texttt{regular} restricts $H$ to
$\{128, 256, 512\}$.
\item[\texttt{dtypes}] a list of dtypes to round-robin across iters.
\item[\texttt{value\_distribution}] how the fuzzer fills tensor
values. \texttt{Uniform} samples $U[-10, 10]$ (default).
\texttt{NaNInjected} replaces 5\% of float elements with NaN,
$\pm$Inf, or 0. \texttt{Adversarial} samples equal-weight across
five dtype-aware buckets: $0$; very small (near the dtype's tiny
value, subnormal for fp32 and fp16); large (max divided by $10$);
wide-uniform $U[-10^3, 10^3]$; and non-finite ($\pm$Inf, NaN).
\end{description}
\par}

\subsection{The seven strategies evaluated}
\label{sec:method:seven}

Table~\ref{tab:strategies} lists the seven strategies. All seven share
the same operator corpus and the same per-op tolerances. Only the
test-input generator changes.

\begin{table}[h]
\centering
\caption{The seven strategies evaluated in the ablation.}
\label{tab:strategies}
\begin{tabular}{llll}
\toprule
strategy & op\_schema & dtypes & value\_distribution \\
\midrule
\texttt{default}          & native             & native     & Uniform        \\
\texttt{boundary}         & $H \in \{1,3,7\}$ & native     & Uniform        \\
\texttt{regular}          & $H \in \{128,256,512\}$ & native & Uniform     \\
\texttt{single\_dtype\_f32} & native           & \{float32\} & Uniform       \\
\texttt{single\_dtype\_f16} & native           & \{float16\} & Uniform       \\
\texttt{nan\_injected}    & native             & native     & NaNInjected    \\
\texttt{adversarial}      & native             & native     & Adversarial    \\
\bottomrule
\end{tabular}
\end{table}

\subsection{Bookkeeping}
\label{sec:method:book}

The driver (\texttt{drivers/p3\_strategies.py}) records per (strategy,
op, iter) the verdict, dtype, shape, layout, \texttt{failure\_kind},
and \texttt{error\_stats}. It also records time-to-first-failure (in
seconds) per (strategy, op) as a wall-clock efficiency metric.

\subsection{Assumptions}
\label{sec:method:assumptions}

The ablation depends on four assumptions.

\begin{enumerate}
\item The 16 controls are correct kernels (human-written Triton or
numpy stand-ins). The companion paper~\cite{gpuemuP1} establishes their
correctness on the gpuemu oracle.
\item The 10 buggy variants are author-seeded with documented LLM
transcription patterns. They are not pulled from real LLM-generated
outputs.
\item Tolerances are fixed per (op, dtype) across all seven strategies.
The companion paper~\cite{gpuemuP2} addresses the tolerance question
separately. Mixing the calibration step in here would conflate two
effects.
\item The Python client decodes received tensors as contiguous, so
layout-only strategies (transposed, strided) are nominal at the client
boundary. The daemon-side fuzzer correctly varies strides but the
kernel sees contiguous data.
\end{enumerate}

\section{Evaluation}
\label{sec:eval}

\textbf{Setup.} RTX~3060, image
\texttt{pytorch/pytorch:2.4.0-cuda12.4-cudnn9-devel}. 7 strategies
$\times$ 26 ops $\times$ 8 iters $= 1{,}456$ cases. Run id
\texttt{run-20260611-101922-4dcac1} on Backblaze B2.

\textbf{Headline: recall and control FP per strategy.}
Each strategy ran the same 10 buggy kernels and the same 16 controls
at 8 iterations per (strategy, kernel). That gives 80 buggy cases
and 128 control cases per strategy. Table~\ref{tab:headline} reports
both axes.

\begin{table}[h]
\centering
\caption{Per-strategy recall (10 buggy $\times$ 8 iters $= 80$) and
control FP rate (16 correct $\times$ 8 = 128).}
\label{tab:headline}
\begin{tabular}{lrr}
\toprule
strategy & bug recall & control FP \\
\midrule
adversarial               & \textbf{79/80 (99\%)}  & \textbf{120/128 (94\%)} \\
nan\_injected             & 75/80 (94\%)           & 93/128 (73\%) \\
\textbf{boundary}         & \textbf{62/80 (78\%)}  & \textbf{0/128 (0\%)} \\
single\_dtype\_f32        & 59/80 (74\%)           & 0/128 (0\%) \\
single\_dtype\_f16        & 59/80 (74\%)           & 3/128 (2\%) \\
default                   & 57/80 (71\%)           & 0/128 (0\%) \\
regular                   & 51/80 (64\%)           & 0/128 (0\%) \\
\bottomrule
\end{tabular}
\end{table}

\begin{figure}[h]
\centering
\includegraphics[width=\textwidth]{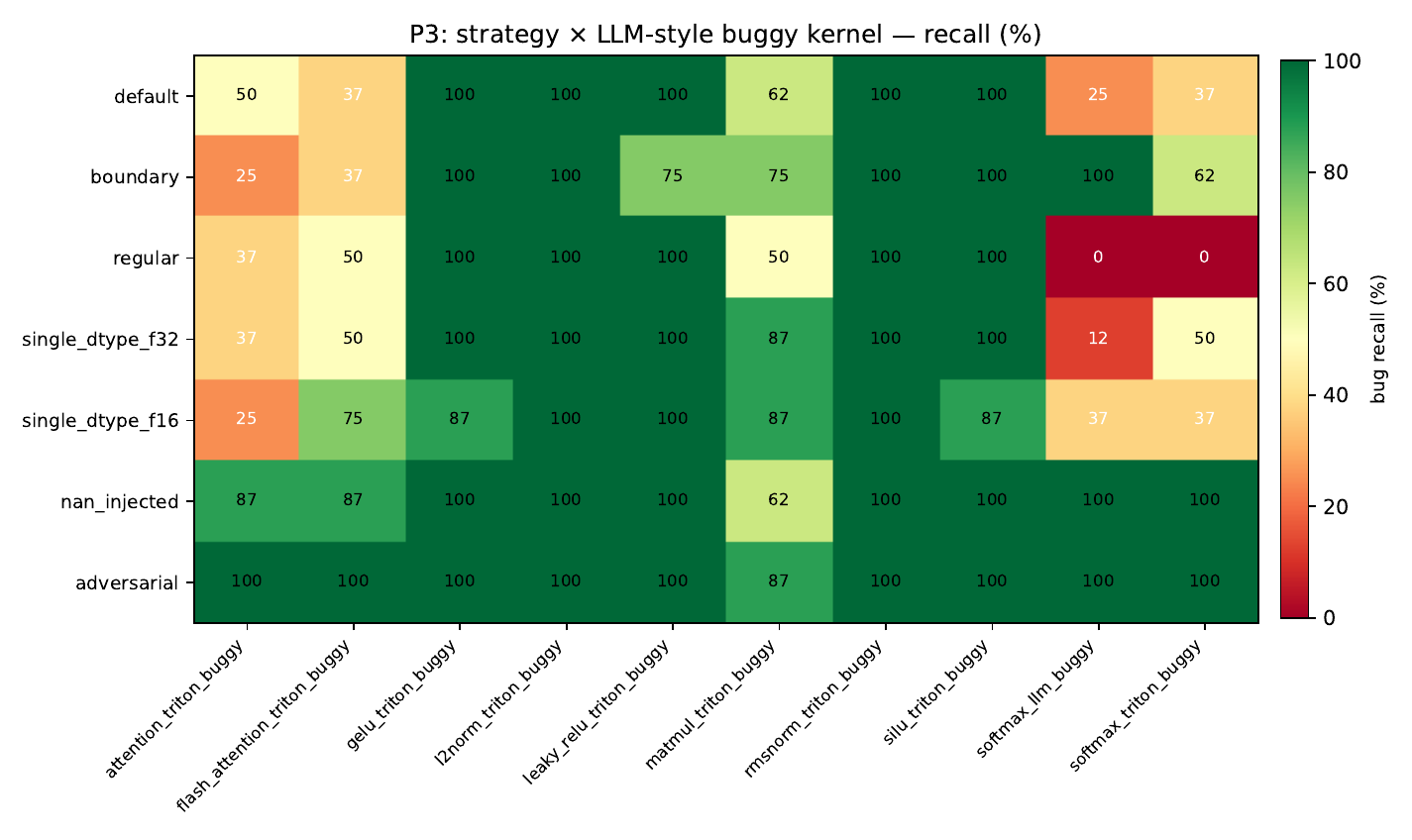}
\caption{Bug recall (\%) per (strategy, buggy kernel), 8 iters
each. Strategy order matches Table~\ref{tab:strategies}
(\texttt{default} top, \texttt{adversarial} bottom).}
\label{fig:heatmap}
\end{figure}

\textbf{Killer finding: shape-dependent bugs vanish under
``regular''.} Table~\ref{tab:shape} reports per-kernel recall on
four shape-sensitive bugs under each of four strategies. Regular
sampling catches 0\% of the two tail-mask bugs. Boundary sampling
catches them at 100\% and 62\% respectively.

\begin{table}[h]
\centering
\caption{Recall on shape-dependent buggy kernels, four strategies,
8 iters per cell.}
\label{tab:shape}
\begin{tabular}{lrrrr}
\toprule
buggy kernel & default & \textbf{boundary} & regular & adversarial \\
\midrule
softmax\_llm\_buggy      & 25\% & \textbf{100\%} & \textbf{0\%} & 100\% \\
softmax\_triton\_buggy   & 38\% & \textbf{62\%}  & \textbf{0\%} & 100\% \\
attention\_triton\_buggy & 50\% & 25\%           & 38\%         & 100\% \\
matmul\_triton\_buggy    & 62\% & 75\%           & 50\%         &  88\% \\
\bottomrule
\end{tabular}
\end{table}

\textbf{Why adversarial and NaN-injected inflate FP.} Both strategies
inject NaN, Inf, and subnormal values into inputs. Correct kernels
that propagate NaN to their output trip the validator's
\texttt{check\_nan} flag, which marks the case as a failure even
though the reference kernel produces the same NaN on the same input.
The 120/128 adversarial FPs and the 93/128 NaN-injected FPs are
therefore validator artefacts of strict NaN handling under non-finite
inputs, not genuine output divergence on correct kernels. The four
strategies that do not inject non-finite inputs (boundary, regular,
default, single\_dtype\_f32) all preserve precision at 0 FP. The
single\_dtype\_f16 strategy registers 3 FPs out of 128, all on the
same fp16-borderline operator. The headline operational
recommendation is therefore boundary as the safe default; adversarial
as a high-recall complement only when the downstream consumer can
itself disregard NaN-output-on-NaN-input failures.

\section{Discussion}
\label{sec:discussion}

The strategy ranking is operator-dependent. The bug families split
into three groups.

\textit{Uniform-magnitude bugs} (gelu missing $0.5$, silu $\beta$
confusion, rmsnorm and l2norm missing \texttt{sqrt}, leaky\_relu
wrong $\alpha$) are caught at 100\% by every strategy. The bug is
shape and value independent, so any input reveals it.

\textit{Shape-dependent tail-mask bugs} (softmax) need the right
boundary shape. Regular sampling at $H \in \{128, 256, 512\}$ (all
powers of two relative to \texttt{BLOCK}) catches 0\% of them.

\textit{Magnitude-sensitive bugs} (attention without $1/\sqrt{D}$
saturates softmax differently at different value scales; matmul
\texttt{acc=} variants behave differently with extreme accumulators)
need value-distribution diversity. Adversarial value sampling has the
highest raw recall on these, but the same value injection that
catches the bug also drives the control FP rate to 94\%.

\textbf{A two-stage operational recipe.} The strategies split
naturally into a gate and a triage pass. Stage one is the gate:
\texttt{boundary} shape sampling on every kernel. It catches 78\% of
seeded bugs and produces zero false alarms on controls, which is what
a CI pipeline needs from a pass/fail signal. Stage two is the triage
pass: \texttt{adversarial} or \texttt{nan\_injected} run on kernels
that the gate already flagged, or on kernels under code review. The
triage pass should not feed back into a pass/fail gate until the
validator stops counting NaN-output as a failure when the reference
kernel produces the same NaN on the same input. In a real testing
pipeline the operator does not know in advance which kernels are
buggy, so the recipe is: gate with boundary, escalate to adversarial
under human review or after the validator change lands.

\section{Limitations}
\label{sec:limitations}

The \texttt{nan\_injected} and \texttt{adversarial} strategies
dominate the raw recall ranking, but the same input distribution
that raises buggy-kernel failure counts also raises control failures
through the validator's strict \texttt{check\_nan} flag. We are not
yet able to separate ``true output divergence on a buggy kernel''
from ``non-finite input propagated to output and flagged by the
validator'' inside the existing fail signal. A validator change that
treats NaN-output as a pass when the reference kernel produces the
same NaN on the same input would split the two and let these
strategies report their bug-discovery recall without the FP
inflation. We treat that as a follow-up in the gpuemu project rather
than re-running the ablation here.

Strategies vary only the candidate set per dim and the dtype and
value distribution. Layout strategies (non-contiguous strided,
transposed) are not yet exercised end to end because the Python
client decodes received tensors as contiguous.

The bug corpus is author-seeded with documented LLM transcription
patterns. We have not yet fuzzed LLM-generated kernels directly.

Iter count per (strategy, kernel) is modest (8). A larger sweep would
tighten confidence intervals on the per-bug recall numbers.

\section{Conclusion}
\label{sec:conclusion}

Test-input generation is not a fixed cost. It is a knob that swings
kernel bug recall by 35 percentage points (64\% under regular shapes
to 99\% under adversarial values). The data argues for one cheap
change and one operationally aware one.

\emph{Always include boundary shapes in the per-op schema.} Without
them, an entire class of tail-mask bugs is invisible to the oracle
(0\% recall on \texttt{softmax\_*\_buggy} under regular sampling, up
to 100\% under boundary sampling). This costs nothing to flip.

\emph{Treat adversarial value sampling as a high-recall complement,
not a default.} Adversarial reaches 99\% raw recall, but on the
current validator it also flags 94\% of correct controls as failures
because non-finite inputs cascade through correct kernels and trip
the NaN check. Either fix the validator to ignore NaN-output when
the reference also produces NaN on the same input, or restrict the
adversarial pass to diagnostic triage on kernels already flagged by
the boundary gate or under code review.

On the gpuemu corpus the boundary knob alone pushes bug recall from
71\% under \texttt{default} to 78\% at 0\% control FP. The
\texttt{adversarial} knob reaches 99\% recall but only becomes a safe
default after the validator stops flagging NaN-output as a failure
when the reference produces the same NaN on the same input. Until
then, treat boundary as the operational gate and adversarial as a
diagnostic pass on suspected-buggy targets.

\paragraph{Artefact.}
The corpus, the strategy driver
(\texttt{drivers/p3\_strategies.py}), the per-strategy table
generator (\texttt{analysis/p3\_strategy\_table.py}), and the replay
tool that fetches the cited run record are bundled in the public
\textsf{gpuemu-corpus} package at
\url{https://github.com/sarkar-dipankar/gpuemu-corpus}. The
validator daemon is at
\url{https://github.com/Skelf-Research/gpuemu}.

\paragraph{License.}
This preprint is released under
\href{https://creativecommons.org/licenses/by/4.0/}{CC-BY 4.0}.

\bibliographystyle{splncs04}
\bibliography{refs}

\end{document}